\documentclass[prb,twocolumn,amsmath,amssymb]{revtex4}
\usepackage{latexsym,amsmath,graphics,graphicx}

\newcommand{\loc}{\mathrm{loc}}
\newcommand{\glob}{\mathrm{glob}}
\newcommand{\bs}{\boldsymbol}
\newcommand{\otop}[1]{\overset{\circ}{#1}\,\!}

\begin{document}
\title{Non-collinear Korringa-Kohn-Rostoker Green function method: 
Application to 3d nanostructures on Ni(001)}

\author{S.~Lounis}\email{s.lounis@fz-juelich.de}
\author{Ph.~Mavropoulos}\email{ph.mavropoulos@fz-juelich.de}
\author{P.~H.~Dederichs}
\author{S.~Bl\"ugel}
\affiliation{Institut f\"ur
Festk\"orperforschung, Forschungszentrum J\"ulich, D-52425 J\"ulich,
Germany}

\date{\today}

\begin{abstract}
Magnetic nanostructures on non-magnetic or magnetic substrates have
attracted strong attention due to the development of new experimental
methods with atomic resolution. Motivated by this progress we have
extended the full-potential Korringa-Kohn-Rostoker (KKR) Green
function method to treat non-collinear magnetic nanostructures on
surfaces. We focus on magnetic 3d impurity nanoclusters, sitting as
adatoms on or in the first surface layer on Ni(001), and investigate
the size and orientation of the local moments and moreover the
stabilization of non-collinear magnetic solutions. While clusters of
Fe, Co, Ni atoms are magnetically collinear, non-collinear magnetic
coupling is expected for Cr and Mn clusters on surfaces of elemental
ferromagnets. The origin of frustration is the competition of the
antiferromagnetic exchange coupling among the Cr or Mn atoms with the
antiferromagnetic (for Cr) or ferromagnetic (for Mn) exchange coupling
between the impurities and the substrate. We find that Cr and Mn
first-neighbouring dimers and a Mn trimer on Ni(001) show
non-collinear behavior nearly degenerate with the most stable
collinear configuration. Increasing the distance between the dimer
atoms leads to a collinear behavior, similar to the one of the single
impurities.  Finally, we compare some of the non-collinear {\it
ab-initio} results to those obtained within a classical Heisenberg
model, where the exchange constants are fitted to total energies of
the collinear states; the agreement is surprisingly good.
\end{abstract}

\maketitle

\section{Introduction}
Theoretically, extensive work is carried out in the area of complex
non-collinear magnetism, particularly for surface and bulk systems. In 
fact a lot of interesting physics would be missed if only collinear 
magnetic structures were considered. In fact,
magnetic nanostructures on magnetic or nonmagnetic substrates are
attractive to the scientific community due to their novel and unusual
properties\cite{wurth,eigler,crommie,monoharan} being of 
relevance both for theory as well as for the applications
in the magnetoelectronics devices.

One of these properties is the non-collinear magnetic order occurring
for geometrically frustrated antiferromagnets, {\it e.g.} on a
triangular lattice, in disordered systems, exchange bias systems, and
molecular magnets, or for systems which exhibit either competing
exchange interactions, or between exchange and spin-orbit
interactions. A simple model for frustration is the following:
Starting with an antiferromagnetic (AF) Cr dimer, the addition a third
Cr atom to form an equilateral triangle leads to a frustrated
geometry. Each atom would like to couple AF to both other atoms.
Since this is impossible, the moments of the three atoms rotate until
a compromise is found. The ground state is then non-collinear,
characterized by an angle of 120$^{\circ}$ between each two atoms. The
same situation will also occur for the AF Cr dimer, since the
interaction of both Cr atoms with the ferromagnetic substrate atoms is
either ferromagnetic of antiferromagnetic. As we will show in this
paper, also in this case a non-collinear structure can result.

The majority of the {\it ab-initio} methods available for the
treatment of non-collinear magnetism make explicit use of Bloch's
theorem and are thus restricted to periodic systems (bulk or
films). Then, even for collinear magnetism, one needs huge supercells
to simulate impurities in a given host (bulk or film) in order to
avoid spurious interactions of the impurities from adjacent
supercells. A few methods have been developed to treat free clusters,
but to our knowledge no {\it ab-initio} methods exist for the
investigation of non-collinear magnetism of clusters in bulk or
deposited on surfaces.

First non-collinear calculations by the KKR Green function method,
though not self--consistent, were already performed in 1985. Oswald
{\it et al.}\cite{oswald0} could show by using the method of
constraints that the exchange interaction between the moments of Mn
and Fe impurity pairs in Cu is in good approximation described by
the $\cos\theta$--dependence of the Heisenberg model.

Sandratskii {\it et al.}\cite{sandratskii1} and K\"ubler {\it et
al.}\cite{kuebler1,kuebler2} pioneered the investigation of
non-collinear magnetic structures using self--consistent density
functional theory. One of the first systems studied by Sandratskii
{\it et al.}\cite{sandratskii1} was the spin spiral of bcc Fe with the
KKR method. Later on, $\Delta$-Fe was a hot topic, and the appearance
of the experimental work of Tsunoda {\it et
al.}\cite{tsunoda1,tsunoda2} led to the development of other
first--principles methods able to deal with non-collinear magnetism
such as LMTO\cite{mryasov}, ASW\cite{knoepfle} and
FLAPW.\cite{kurz,nordstroem,sjoestedt}

Several papers\cite{sandratskii2,sandratskii3} describe how symmetry
simplifies the calculational effort for the spiral magnetic structures
in the case of perfect periodic systems---this involves the
generalized Bloch theorem. In {\it ab-initio} methods, this principle
is used together with the constrained density functional
theory\cite{dederichs,grotheer} giving the opportunity of studying
arbitrary magnetic configurations where the orientations of the local
moments are constrained to nonequilibrium directions.

Concerning free clusters, few methods are developed. For example, Oda
{\it et al.}\cite{oda} developed a plane-wave pseudopotential scheme
for non-collinear magnetic structures. They applied it to small Fe
clusters for which they found non-collinear magnetic structures for
Fe$_{5}$ and linear-shape Fe$_{3}$. This last result was in
contradiction with the work of Hobbs {\it et al.}\cite{hobbs} who
found only a collinear ferromagnetic configuration using a projector
augmented-wave method. Small Cr clusters were found magnetically
non-collinear,\cite{oda} as shown also by Kohl and Bertsch\cite{kohl}
using a relativistic nonlocal pseudopotential method after
optimization of the ionic structure by a Monte Carlo technique.
However, within the generalized gradient approximation (GGA) of
density-functional theory, Hobbs {\it et al.}\cite{hobbs} find that in
many cases the non-collinear states can be metastable, while the
ground-state solutions are collinear and arise after geometrical
optimization of the free-standing clusters.

One main result of Oda {\it et al.}\cite{oda} and Hobbs {\it et
al.}\cite{hobbs} concerns the variation of the magnetization density
with the position. The spin direction changes in the interstitial
region between the atoms where the charge and magnetization densities
are small, while the magnetization is practically collinear within the
atomic spheres. This supports the use of a single spin direction for
each atomic sphere as an approximation in order to accelerate the
computation; this approximation is followed also here.

The aim of this work is to present a method  based on the
full-potential KKR scheme\cite{papanikolaou} which can deal with
non-collinear magnetism in systems of reduced symmetry. This method is 
ideal for treating impurities or small clusters on surfaces or in bulk.
As an application we study small 3d clusters on the Ni(001) surface where we
find complex magnetic configurations.

\section{Non-collinear KKR formalism}
The KKR method uses multiple-scattering theory in order to determine
the one-electron Green function in a mixed site and angular--momentum
representation. The retarded Green function is expanded as:
\begin{widetext}
\begin{equation}
{G}(\vec{R}_n+\vec{r},\vec{R}_{n'}+\vec{r}';E) = 
-i \sqrt{E} \sum_{L}{R}_{L}^n({\vec r_<};E){H}_{L}^n({\vec r_>};E)\delta_{nn'}
+\sum_{LL'}{R}_{L}^n({\vec r};E) {G}_{LL'}^{nn'}(E)
{R}_{L'}^{n'}({\vec r'};E)
\label{eq:1}
\end{equation}
\end{widetext}
Here, $E$ is the energy and $\vec{R}_{n}$, $\vec{R}_{n'}$ refer
to the atomic nuclei positions. By $\vec{r}_<$ and $\vec{r}_>$ we
denote respectively the shorter and longer of the vectors $\vec{r}$
and $\vec{r}'$ which define the position in each Wigner--Seitz cell
relative to the position $\vec{R}_{n}$ or $\vec{R}_{n'}$.  
The wavefunctions ${R}_{L}^n(\vec{r};E)$ and
${H}_{L}^n(\vec{r};E)$ are, respectively, the regular and irregular
solutions of the Schr\"odinger equation for the potential $V_n$ at
site $n$, being embedded in free space; $L=(l,m)$ is a combined index
for angular momentum quantum numbers; $l$ is truncated at a maximum
value of $l_{\mathrm{max}}$. The first term on the LHS of
Equation~(\ref{eq:1}) is the so-called \emph{single site scattering
term}, which describes the behavior of an atom $n$ in free space. All
multiple- and back-scattering information is contained in the second
\emph{back-scattering} term via the structural Green functions
${G}_{LL'}^{nn'}(E)$ which are obtained by solving the algebraic Dyson
equation:
\begin{eqnarray}
{G}_{LL'}^{nn'}(E) &=& \otop{G}^{nn'}_{LL'}(E) \nonumber\\
&+&\!\!\!\!\! \sum_{n'',L''L'''}\!\!\!\!\! \otop{G}_{LL''}^{nn''}(E) 
{\Delta t}_{L''L'''}^{n''}(E) {G}_{L'''L'}^{n''n'}(E)
\label{eq:2}
\end{eqnarray}
Equation~(\ref{eq:2}) follows directly from the usual Dyson eq.~of the
form $G=\otop{G}+\otop{G}\,\Delta V\,G$, with $\Delta V$ the 
perturbation in the
potential and $\otop{G}$ the reference system Green function. The 
summation in (\ref{eq:2}) is over all lattice sites
$n''$ and angular momenta $L''$ for which the perturbation ${\Delta
t}_{L''L'''}^{n''}(E)={t}_{L''L'''}^{n''}(E) -\otop{t}_{L''L'''}^{n''}(E)
$ between the ${t}$ matrices of the real and the reference system is
significant (the $t$-matrix gives the scattering amplitude of the
atomic potential). The quantities $\otop{G}_{LL'}^{nn'}(E)$ are the
structural Green functions of the reference system. For the
calculation of a crystal bulk or surface, the reference system can be
free space, or, within the tight-binding KKR formulation,\cite{SKKR} a
system of periodically arrayed repulsive potentials. After the host
(bulk or surface) Green function is found, it can be used in a second
step as a reference for the calculation of the Green function of an
impurity or a cluster of impurities embedded in the host.

The algebraic Dyson equation (\ref{eq:2}) is solved by 
matrix inversion, as we will
see later on in Equation~(\ref{eq:15}). In case of spin-dependent
electronic structure, spin indexes enter in the $t$-matrix, the Green
functions and in Eq.~(\ref{eq:2}). Especially in the case of
non-collinear magnetism, these quantities become $2\times 2$ matrices
in spin space, denoted by $\bs{t}$ and $\bs{G}$.

Once the spin-dependent Green function is known, all physical
properties can be derived from it. In particular, the charge density
$n(\vec{r})$ and spin density $\vec{m}(\vec{r})$ are given
by an integration of the imaginary part of $\bs{G}$ up to the Fermi level
$E_F$ and a trace over spin indexes $s$ (putting the Green
function in a matrix form in spin space):
\begin{eqnarray}
n(\vec{r})&=&
-\frac{1}{\pi}\mathrm{Im}\mathrm{Tr}_{s}
\int^{E_F}\bs{G}(\vec{r},\vec{r};E)\,dE
\label{eq:2.5} \\
\vec{m}(\vec{r})&=&
-\frac{1}{\pi}\mathrm{Im}\mathrm{Tr}_{s}
\int^{E_F}\vec{\bs{\sigma}}\,\bs{G}(\vec{r},\vec{r};E)\,dE.
\label{eq:2.6} 
\end{eqnarray}
Here, $\vec{\bs{\sigma}}=(\bs{\sigma}_x,\bs{\sigma}_y,\bs{\sigma}_z)$
are the Pauli matrices and $\mathrm{Tr}_{s}$ means the trace operation 
in spin space.

The basic difference between non-collinear and collinear magnetism is
the absence of a natural spin quantization axis common to the whole
crystal. The density matrix is not anymore diagonal in spin space as
in the case of collinear magnetism. Instead, in any fixed frame of
reference it has the form
\begin{equation}
\bs{\rho}(\vec{r})=
\begin{bmatrix}
{\rho}_{\uparrow\uparrow}(\vec{r}) & {\rho}_{\uparrow\downarrow}(\vec{r}) \\
{\rho}_{\downarrow\uparrow}(\vec{r}) & {\rho}_{\downarrow\downarrow}(\vec{r})
\end{bmatrix}
= 
\frac{1}{2}\left[n(\vec{r}) + \vec{\bs{\sigma}} \cdot  \vec{m}(\vec{r})\right]
\label{eq:3}
\end{equation}
At any \emph{particular point} in space, of course, a \emph{local}
frame of reference can be found in which $\bs{\rho}$ is diagonal, but
this local frame can change from point to point.

In order to deal with non-collinear magnetism, we have to solve the
appropriate Dyson equation. First we define the reference system which
is a perfect surface characterized by collinear magnetism. Although
the collinearity of the reference system is not a necessary
requirement, it serves our purpose of calculating the electronic structure 
of the ferromagnetic or nonmagnetic surfaces which are used as 
reference systems. Thus the host Green functions $\otop{\bs{G}}$ 
and $t$-matrices $\otop{\bs{t}}$ are assumed diagonal in spin space. In this way, 
in the case of a magnetic host, a global spin frame of reference 
is defined. The host $\otop{\bs{G}}$ and $\otop{\bs{t}}$ are thus of the form:
\begin{equation}
\!\!\!\!\!\!\! \otop{\bs{G}}(E)=
\begin{bmatrix}
\otop{G}_{\uparrow\uparrow}(E) & 0 \\
0 & \otop{G}_{\downarrow\downarrow}(E)
\end{bmatrix};
\otop{\bs{t}}(E)=
\begin{bmatrix}
\otop{t}_{\uparrow\uparrow}(E) & 0 \\
0 & \otop{t}_{\downarrow\downarrow}(E)
\end{bmatrix}
\label{eq:4}
\end{equation}
Then the perturbed system is constructed. The impurity atoms which
might couple magnetically in a non-collinear way reside on the
surface, perturbing the potential at a few neighboring sites (atoms or
empty cells representing the potential in the vacuum). Within this finite 
cluster of perturbed sites the magnetization can be non-collinear leading to the
appearance of non-diagonal elements of the $t$-matrix:
\begin{equation}
\bs{t}(E)= 
\begin{bmatrix}
{t}_{\uparrow\uparrow}(E) & {t}_{\uparrow\downarrow}(E)\\
{t}_{\downarrow\uparrow}(E) & {t}_{\downarrow\downarrow}(E)
\end{bmatrix}
\label{eq:6}
\end{equation}
The non-diagonal $t$-matrix contains the information on spin-flip scattering
by the atomic potential.

At this stage an approximation enters our method. It is assumed that,
separately for each atom, there exists an intra-atomic spin
quantization axis common to the whole atomic cell. This axis is
identified with the spatial average of the magnetization density
${\vec{m}_n(\vec{r})}$ in each cell $n$. This defines the local spin
frame of reference. In this way we neglect the variation of the spin
quantization axis within the cell during self consistency, avoiding
the time-consuming numerical solution of the potential of coupled
Schr\"odinger equations of the two spin channels. Within the local
density approximation of density-functional theory, the exchange
correlation potential has the same reference frame as the local
magnetization ${\vec{m}_n(\vec{r})}$. Then for each atom we have a
potential which is collinear in the local frame, and the solutions of
the Schr\"odinger equation, ${R}^{\loc}_{nLs}(\vec{r};E)$ and
${H}^{\loc}_{nLs}(\vec{r};E)$, depending on the spin index $s$ of the
local frame.

The solution of the Schr\"odinger equation separately for each spin
channel provides also the diagonal $t$-matrix of each atomic 
cell $n$ in the local frame of reference:
\begin{equation}
\bs{t}_n^{\loc}(E) =  
\begin{bmatrix}
t^{\loc}_{\uparrow\uparrow}(E) & 0 \\
0 & t^{\loc}_{\downarrow\downarrow}(E)
\end{bmatrix}
\label{eq:7}
\end{equation}
Then the $t$-matrix is rotated from the local to the global spin frame
of reference using the spin rotation matrix $\bs{U}_n$:
\begin{eqnarray}
\bs{t}_n^{\glob}(E)&=& \bs{U}_n \bs{t}_n^{\loc}(E) \bs{U}_n^{\dag},
\label{eq:8}
\end{eqnarray}
$\bs{U}_n$ being given by
\begin{eqnarray}
\bs{U}_n &=& \begin{bmatrix}
{\cos(\frac{\theta_n}{2}) e^{-\frac{i}{2}\phi_n}}&
{-\sin(\frac{\theta_n}{2}) e^{-\frac{i}{2}\phi_n}} \\
{\sin(\frac{\theta_n}{2}) e^{\frac{i}{2}\phi_n}}&
{\cos(\frac{\theta_n}{2}) e^{\frac{i}{2}\phi_n}} 
\end{bmatrix} .
\label{eq:9}
\end{eqnarray}
The polar angles ${\theta_n}$ and ${\phi_n}$ define the direction of
the local magnetic moment with respect to the global spin frame of
reference. Normally, $\theta_n$ and $\phi_n$ vary within the atomic
cell, but in the approximation used here, average angles are defined
for each cell via an averaging of the magnetization density within the
cell. Of course, when self-consistency is achieved, both the averaged
and the point-by-point varying magnetization direction can be extracted 
from the output density matrix. Thus the assumption of a unique spin 
direction in each cell is only made for the spin-dependent potential.

The $t$-matrix in the global spin frame of reference can be rewritten
in the following way:
\begin{widetext}
\begin{equation}
\bs{t}_n^{\glob}(E)=\bs{U}_n \left[ \frac{1}{2}({t}_{\uparrow\uparrow}^{\loc}(E) + 
{t}_{\downarrow\downarrow}^{\loc}(E)) \bs{1} + 
\frac{1}{2}({t}_{\uparrow\uparrow}^{\loc}(E) -
{t}_{\downarrow\downarrow}^{\loc}(E))\bs{\sigma}_z \right] \bs{U}_n^{\dag}
\label{eq:10}
\end{equation}
\end{widetext}
with $\bs{\sigma}_z$ is the $z$ component of the Pauli matrices:
\begin{equation}
\bs{\sigma}_{z}=
\begin{bmatrix}
1 &  0\\
0 & -1
\end{bmatrix}
\label{eq:11}
\end{equation}

It is convenient to define the projection matrices $\bs{\sigma}_{ns}$
for the local spin--up ($\uparrow$) and spin--down ($\downarrow$)
directions as:
\begin{equation}
\bs{\sigma}_{ns} = \frac{1}{2} \bs{U}_n 
(\bs{1} \pm \bs{\sigma}_z) \bs{U}^{\dag}_n = (\bs{\sigma}_{ns})^2
\ \ \ (\mbox{$+$ for $s=\uparrow$, $-$ for $s=\downarrow$})
\label{eq:12}
\end{equation}
Then $\bs{t}^{\glob}_n(E)$ is written as:
\begin{equation}
\bs{t}^{\glob}_n(E)=t_{n\uparrow\uparrow}^{\loc}(E)\bs{\sigma}_{n\uparrow} 
+ t_{n\downarrow\downarrow}^{\loc}(E)\bs{\sigma}_{n\downarrow} 
\label{eq:13}
\end{equation}
In the collinear case the local and global frames are identical and
the projection operators reduce to:
\begin{equation}
\bs{\sigma}_{\uparrow}=
\begin{bmatrix}
{1} & {0}\\
{0} & {0}
\end{bmatrix},\ \ \ 
\bs{\sigma}_{\downarrow}=
\begin{bmatrix}
{0} & {0}\\
{0} & {1}
\end{bmatrix} \ \ \ \mbox{(collinear case)}.
\label{eq:14}
\end{equation}
At this stage, the difference between the $t$-matrices $\Delta
\bs{t}_n^{\glob}=\bs{t}_n^{\glob}-\otop{\bs{t}}_n$ is calculated in order
to get all the ingredients to solve the Dyson equation for the
structural Green function ($\otop{\bs{t}}_n$ has been defined in the global 
frame in Eq.~(\ref{eq:4})). This is the analogue of Eq.~(\ref{eq:2})
in matrix form in spin space:
\begin{equation}
\bs{G}_{\mathrm{str}}(E) =
\otop{\bs{G}}_{\mathrm{str}}(E) + \otop{\bs{G}}_{\mathrm{str}}(E) \Delta \bs{t}^{\mathrm{\glob}}(E)
{\bs{G}}_{\mathrm{str}}(E).
\label{eq:15a}
\end{equation}
Here, in analogy to eq.~(\ref{eq:1}) and eq.~(\ref{eq:2}),
$\bs{G}_\mathrm{str}(E)$ are matrices of size $2\times 2$ in spin space,
size $(l_{\mathrm{max}}+1)^2 \times (l_{\mathrm{max}}+1)^2$ in angular
momentum space, and size $N\times N$ (with $N$ the number of sites) in
real space; all these indices are combined to form $2\times
(l_{\mathrm{max}}+1)^2\times N$-dimensional matrices. The $t$-matrix
itself is diagonal in real space site indexes.  The solution of
Eq.~(\ref{eq:15a}) for the structural Green function requires matrix
inversion, yielding $\bs{G}_\mathrm{str}(E)$ in the global frame:
\begin{equation}
\bs{G}_{\mathrm{str}}(E) =
\otop{\bs{G}}_{\mathrm{str}}(E)(1 - \Delta \bs{t}^{\mathrm{\glob}}(E)\, 
\otop{\bs{G}}_{\mathrm{str}}(E))^{-1}.
\label{eq:15}
\end{equation}

Equation~(\ref{eq:1}) can be now rewritten in the non-collinear case in
order to obtain the Green function in the global frame. Using the
matrices $\bs{\sigma}_{ns}$ (Eq.~(\ref{eq:12})) to project the local
wavefunctions to the global frame, the Green function is written as:
\begin{widetext}
\begin{eqnarray}
\bs{G}^{\glob}(\vec{R}_n+{\vec r},\vec{R}_{n'}+{\vec r'};E) =
-i \sqrt{E}\sum_{Ls}
{R}^{\loc}_{nLs}({\vec r_<};E)
{H}^{\loc}_{nLs}({\vec r_>};E)
\bs{\sigma}_{ns}
\nonumber\\
+\sum_{LL'ss'}
{R}^{\loc}_{nLs}({\vec r};E) 
\bs{\sigma}_{ns} 
\bs{G}^{\glob}_{LL'nn'}(E)
\bs{\sigma}_{n's'} 
{R}^{\loc}_{n'L's'}({\vec r'};E).
\label{eq:20}
\end{eqnarray}
\end{widetext}
If needed, the Green function can be rotated to the local frame of any
atom by the use of the transformation matrices $\bs{U}_n$
(Eq.~(\ref{eq:9})).  We point out that, even in the local frame of
reference, the Green function is not in general diagonal in spin
space. Finally we calculate the charge density and spin density from
Equations~(\ref{eq:2.5},\ref{eq:2.6}). The spin dependent local
density of states within the Wigner-Seitz cell WS of each site $n$ is:
\begin{equation}
{n}_{ns}(E) = -\frac{1}{\pi}\int_{\mathrm{WS}} 
\mathrm{Im} {G}_{ss}(\vec{R}_n+\vec{r},\vec{R}_n+\vec{r};E) d^{3}r
\label{eq:22}
\end{equation}
The spin density $\vec{m}=(m^x,m^y,m^z)$
(Eq.(\ref{eq:2.6})) is non-collinear.  The new polar angles at each
site $n$ can then be obtained for each point by
\begin{equation}
{\tan} {\theta}_{n}({\vec r}) = \frac{{m}^z_{n}({\vec r})}
{{m}_{n}({\vec r})} 
, \ \ \ 
{\tan} \phi_{n}(\vec{r}) = \frac{{m}^y_{n}(\vec{r})}
{{m}^x_{n}(\vec{r})}
\label{eq:24}
\end{equation}
or as an average over the local Wigner--Seitz cell
\begin{equation}
{\tan} {\theta}_{n}= \frac
{\int_{\mathrm{WS}}{m}^z_{n}({\vec r}) d{\vec r}}
{\int_{\mathrm{WS}}{m}_{n}({\vec r}) d{\vec r} } 
, \ \ \ 
{\tan} {\phi}_{n} = \frac
{\int_{\mathrm{WS}}{m}^y_{n}({\vec r}) d{\vec r}}
{\int_{\mathrm{WS}}{m}^x_{n}({\vec r}) d{\vec r}} 
.
\label{eq:25}
\end{equation}

\section{Applications}
As an application of our method, we study the magnetic state of 3{\it d}--atom 
clusters in and on the Ni(001) surface. In a first step, we study the adatom 
properties, which are already known from previous work. In a second step, we 
perform calculations for 3{\it d} dimers and trimers and use the understanding 
gained from the single adatoms in order to explain the results.

Our calculations henceforth are based on the Local Spin Density
Approximation (LSDA) of density functional theory with the
parametrization of Vosko {\it et al.}\cite{vosko}. The full
nonspherical potential was used, taking into account the correct
description of the Wigner--Seitz atomic cells.\cite{Stefanou90}
Angular momenta up to $l_{\mathrm{max}} = 3$ were included in the
expansion of the Green functions and up to $2l_{\mathrm{max}} = 6$ in
the charge density expansion. Relativistic effects were described in
the scalar relativistic approximation.

First, the surface Green functions are determined by the screened KKR
method\cite{SKKR} for the (001) surface of Ni which serves as the
reference system. The equilibrium lattice parameter of Ni was used
(6.46~{a.u.}  $\approx$~3.42~{\AA}). To describe the impurities on the
surface (later we refer to these as {\it adatoms} and to the
impurities sitting in the first surface layer as {\it inatoms}), we
consider a cluster of perturbed potentials which includes the
potentials of the impurities and the perturbed potentials of several
neighboring shells, with typical size ranging from 19 perturbed sites
for the single impurity to 32 for the dimers and trimers; in all
cases, at least the first neighboring sites of the impurity atoms were
taken into account in the calculation to ensure the correct screening of the
impurity potential. Test calculations have shown that this is adequate 
for our work; this is a merit of the Green function method, in which
the correct boundary conditions of the host (in our case of the host surface) 
are included in the Green function via the Dyson equation. We consider 
the adatoms at the unrelaxed hollow position in the first vacuum layer, and 
the inatoms at the unrelaxed position in the first surface layer.

The orientations assigned to the spin moments of the impurities are
always relative to the orientation of the substrate moment, which we
take as the global frame. This, in turn, depends on delicate physical
quantities such as the magnetic anisotropy energy, which cannot be 
related to the local properties of the small clusters
that we study.  In the present approach such effects arising from spin-orbit 
interaction are not included. The direction of the host moments must therefore be
considered as an input parameter from experiments or from independent
{\it ab-initio} calculations.

\subsection{3{\it d} single adatoms and inatoms}
3{\it d} adatoms on Fe(001) and on Ni(001) have been already studied
previously, using the KKR method\cite{nonas,nonas1,nonas2,nonas3} in
the atomic sphere approximation.  Here we repeat the calculations of
3d adatoms on Ni(001) using the full potential method (a detailed 
work on Fe and Co on Ni(001) is presented in a recent 
article\cite{mavropoulos}). We 
give a brief analysis of the results, which are basically 
unchanged, in order to use them as a step for understanding 
the behavior of dimers and trimers later on.

A collinear calculation of the magnetic state of a single adatom on a
ferromagnetic substrate can give in some cases two solutions: one with
ferromagnetic coupling (FM) to the substrate and one with
antiferromagnetic coupling (AF). One of these states will correspond
to the real ground state, and the other to a local minimum; this is
actually a local minimum with respect to collinear variations of the
magnetic moment, since the angle $\theta$ between the local moment and
the substrate moment cannot be varied in a collinear calculation. From
total energy calculations of the two states, the ground state can be then
determined. In some cases, when the intra-atomic exchange field is
not strong (beginning or end of the 3d series), only one of the two
minima exists. On the other hand, if non-collinear effects are included
in the calculations, one of the two minima usually becomes unstable
against an angular rotation of the moment, {\it i.e.}, it is then
actually a saddle point.

\begin{figure}[h!]
\begin{center}
\includegraphics*[width=1.0\linewidth]{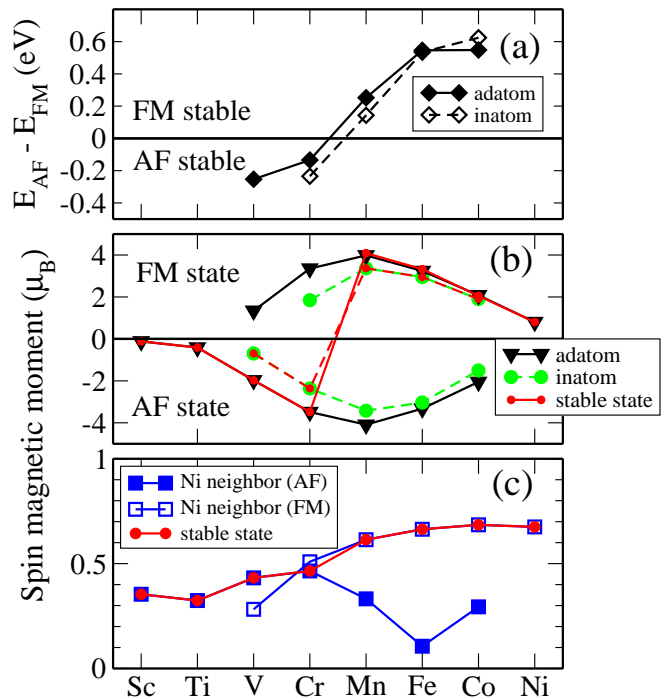}
\caption{3d adatoms and inatoms on Ni(001): (a) Energy difference between the
AF and FM coupling, the values related to adatom and inatoms are described by 
respectively full and empty black diamonds; (b) 
magnetic moments of the adatoms (black triangles) and inatoms (green circles) 
within the 2 possible magnetic configurations FM and AF; (c) the variation of the 
magnetic moments of Ni first Nearest Neighbors of the adatoms.}
\label{adatom-energy}
\end{center}
\end{figure}

\begin{figure*}[!ht]
\begin{center}
\includegraphics*[angle=270,width=\linewidth]{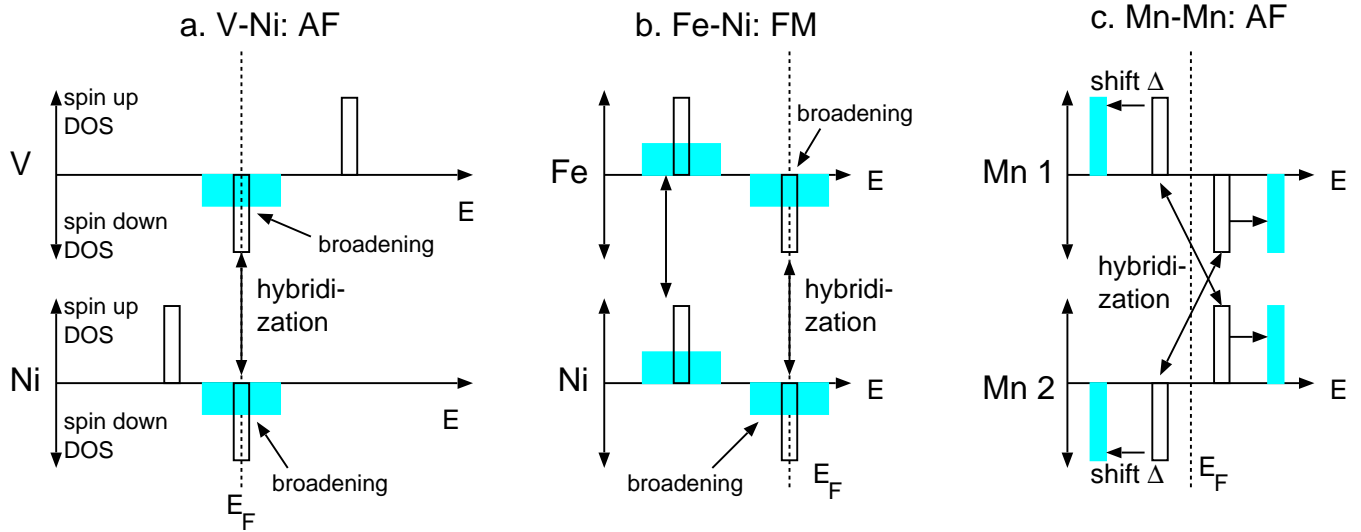}
\caption{Alexander--Anderson model for neighboring magnetic atoms: (a) Early 
3{\it d} transition elements in interaction with Ni surface atoms; (b) Late 
3{\it d} transition elements in interaction with Ni surface atoms; (c) Cr or Mn dimer.}
\label{alexander-anderson}
\end{center}
\end{figure*}

The full diamonds in the Fig.~\ref{adatom-energy}(a) show the energy
difference between the AF and the FM solution for 3d adatoms on
Ni(001). The first elements of the 3d series (Sc, Ti, V, Cr) are AF
coupled to the substrate whereas the coupling of Mn, Fe, Co and Ni is
FM. Sc (AF), Ti (AF), and Ni (FM) are characterized by a single
solution. Clearly, the AF-FM transition occurs when the adatom atomic
number changes from Cr ($Z=24$) to Mn ($Z=25$). This transition can be
interpreted as in the case of the interatomic interaction of magnetic
dimers,\cite{anderson,oswald} in terms of the energy gain due to the
formation of hybrid states with the Ni substrate as the $3d$ virtual
bound state comes lower in energy with increasing $Z$. An explanation
(see Fig.\ref{alexander-anderson}) can be given in terms of the {\it
d--d} hybridization between the adatom 3{\it d} states and the Ni
substrate 3{\it d} states. Energy is gained when a half--occupied {\it
d} virtual bound state (VBS) at $E_F$ is broadened by 
hybridization with the Ni
minority 3{\it d} states, which lie at $E_F$ (the Ni majority {\it d}
states are fully occupied and positioned below $E_F$). For the early
3{\it d} adatoms (Fig.\ref{alexander-anderson}a), it is the majority
{\it d} VBS which is at $E_F$, thus the majority--spin direction of
the adatom is favourably aligned with the minority--spin direction of
Ni, and an AF coupling arises. For the late 3{\it d} adatoms
(Fig.\ref{alexander-anderson}b), on the contrary, the minority {\it
d} VBS is at $E_F$, and this aligns with the Ni minority {\it d}
states; then a FM coupling arises. For our purposes we keep in mind
that, since Cr and Mn are in the intermediate region, {\it i.e.}, near
the AF-FM transition point, their magnetic coupling to the Ni
substrate is weak; this has consequences to be seen in the behavior of
dimers, trimers, etc., in the next subsections.

The magnetic moments of the adatoms and Ni first neighbors in the
surface layer are shown in Fig.~\ref{adatom-energy}(b) and
(c). Evidently the moment of the Ni first neighbors is strongly
affected by the adatoms. Especially in the AF state for Mn, Fe, and Co
adatoms, the Ni moment is strongly reduced and the FM configuration 
is stable. As regards the adatom
moments, due to its half filled d band the Mn adatom carries the
highest magnetic moment (4.09 $\mu_B$) followed by Cr (3.48 $\mu_B$)
and Fe (3.24$\mu_B$). 

To understand the effect of coordination and stronger hybridization on
the magnetic behavior of the adatoms, we take the case of impurities
sitting in the first surface layer (inatoms). We carried out the
calculations for V, Cr, Mn, Fe and Co impurities. The corresponding
spin moments are shown in Fig.\ref{adatom-energy}b (green circles),
and the FM-AF energy differences are shown in
Fig.~\ref{adatom-energy}a (open diamonds and dashed line).

Compared to the adatom case, the spin moments are reduced, especially
for V and Cr. This effect is expected due to the increase of the
coordination number from 4 to 8 and the subsequent stronger
hybridization of the 3{\it d} levels with the host
wavefunctions. Moreover, the energy difference $\Delta E$ between the
AF and FM solutions is affected. The trend can be understood as
follows. In the case of Cr, the reduction of the local magnetic moment
M is accompanied by a reduction of the exchange splitting $\Delta E_X$
as $\Delta E_X \approx I \cdot M$, where $I \approx$~1eV is the
intra-atomic exchange integral.  This means that, for the inatom, the
occupied 3{\it d} states are closer to $E_F$ than for the adatom. In
turn, this intensifies the hybridization of these states with the Ni
3{\it d} states (which are close to $E_F$). At the same time, also the
higher coordination number intensifies the {\it d-d} hybridization.
The hybridization-induced level shift in the AF configuration
increases, and the energy of the AF state is thus lowered. The same
mechanism is responsible for the weakening of the FM coupling of Mn
inatom compared to the adatom. Similarly, the stronger hybridization
of the Co--inatom {\it d}--states stabilizes even more its FM
configuration due to the energy gain from the broadening of the {\it
d} virtual bound state.

\subsection{Adatom and inatom dimers}
Having established the single adatom behavior, we turn to adatom
dimers. We considered three geometries of increasing distance: dimers
as first, second, and fourth neighbors. We will discuss the magnetic
interaction between the dimer atoms and the resulting magnetic order,
first looking only at collinear states and then allowing for
non-collinear order. We will see how, in certain cases, the collinear
state reduces the symmetry, while the non-collinear state restores the
full symmetry of the system. Non-collinear order is finally established 
for certain first-neighbor dimers.

\begin{figure}[!h]
\begin{center}
\includegraphics*[width=1.0\linewidth]{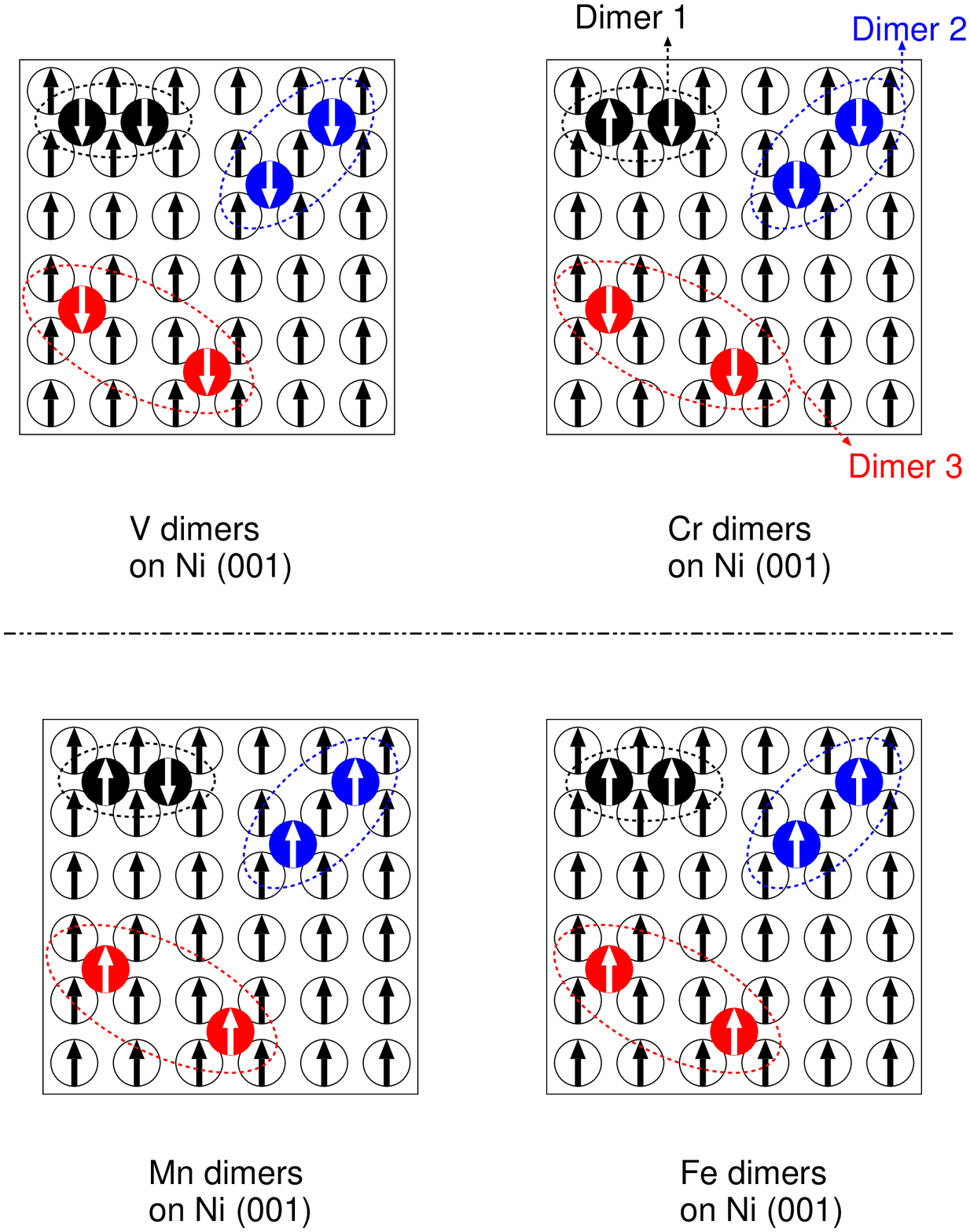}
\caption{Different geometrical configurations considered for dimers at
the surface of Ni(001). Dimer-1--type corresponds to the case where the
atoms are first neighboring atoms, dimer-2--type where the atoms are
2'NN and finally dimer-3--type to 4'NN. The collinear magnetic ground
state are also shown for V, Cr, Mn and Fe dimers.}
\label{geometry-dimer}
\end{center}
\end{figure}

Fig.~\ref{geometry-dimer} represents schematically the different
considered geometrical configurations of impurity dimers residing on
the surface. We have investigated the dimer-1--type of geometry (the
adatoms are first neighboring atoms), dimer-2--type (the adatoms are
second neighbors) and dimer-3--type (the adatoms are fourth
neighbors). This allows us to monitor the strength of the magnetic
coupling as a function of the distance. Three collinear magnetic
configurations were treated: (i) antiferromagnetic coupling within the
dimer leading to a ferrimagnetic solution (Ferri), (ii) ferromagnetic
coupling within the dimer with both atoms ferromagnetically coupled to
the substrate (FM), or (iii) ferromagnetic coupling within the dimer
with both atoms antiferromagnetically coupled to the substrate (AF).

Our calculations include V, Cr, Mn and Fe dimers. We found that all V
and Fe dimer types behave like the adatoms: in all geometries, both V
atoms are AF and both Fe atoms are FM. On the other hand, Cr and Mn
dimers show magnetic frustration. As shown in
Fig.~\ref{geometry-dimer}, both the Cr--dimer-1 and Mn--dimer-1 show (in
a collinear calculation) a Ferri ground state (see Table~\ref{energy-dimer}).
 With increasing distance between the adatoms, a transition occurs 
to the single 
adatom magnetic behavior which is AF for Cr--dimers and FM for
Mn--dimers. It is clear that, in the dimer-1 case, there is a
competition of exchange interactions.

\begin{figure}[h!]
\begin{center}
(a)
\includegraphics*[width=0.4\linewidth]{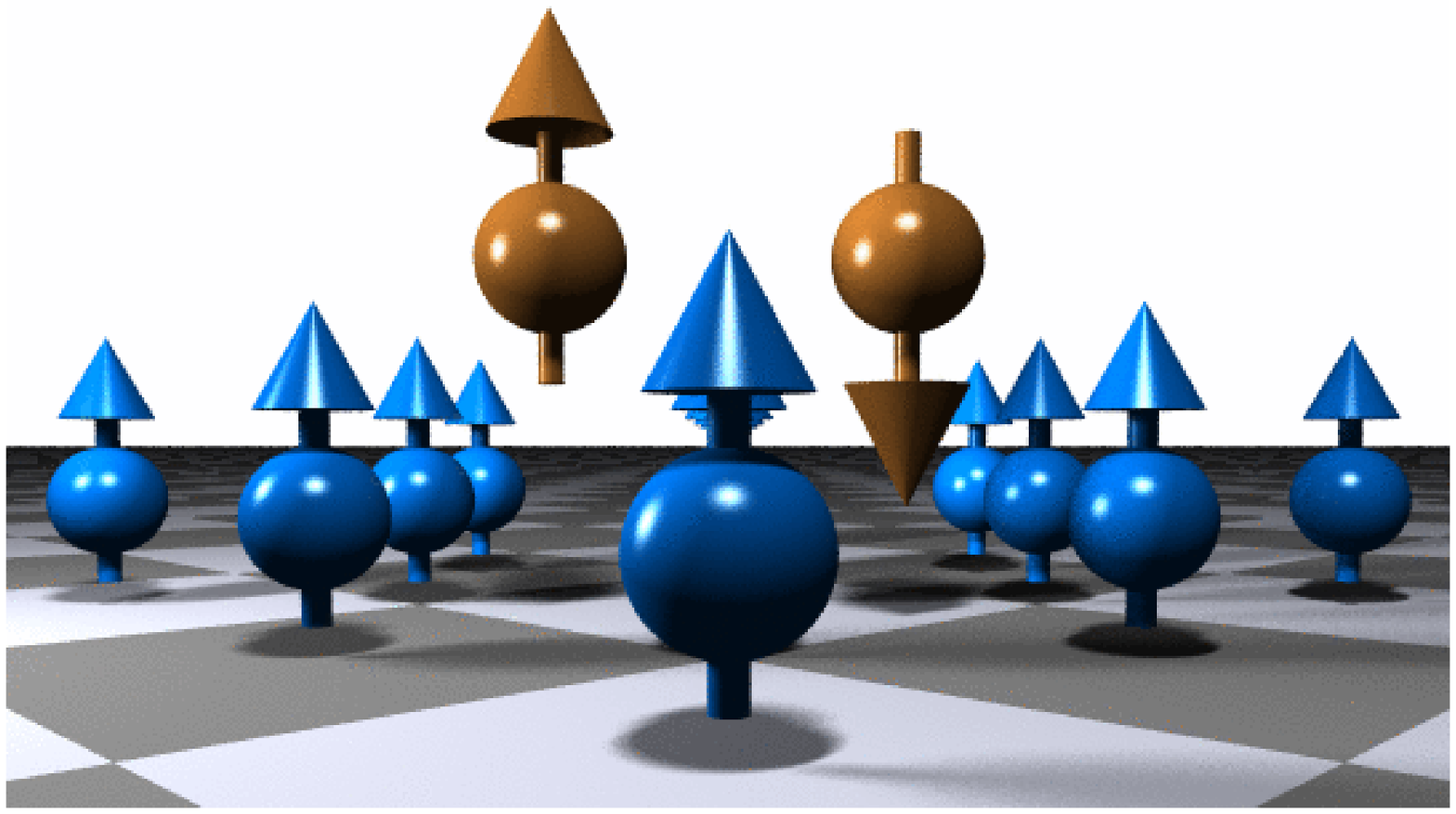}
\hspace{-0.1cm}
(b)
\includegraphics*[width=0.4\linewidth]{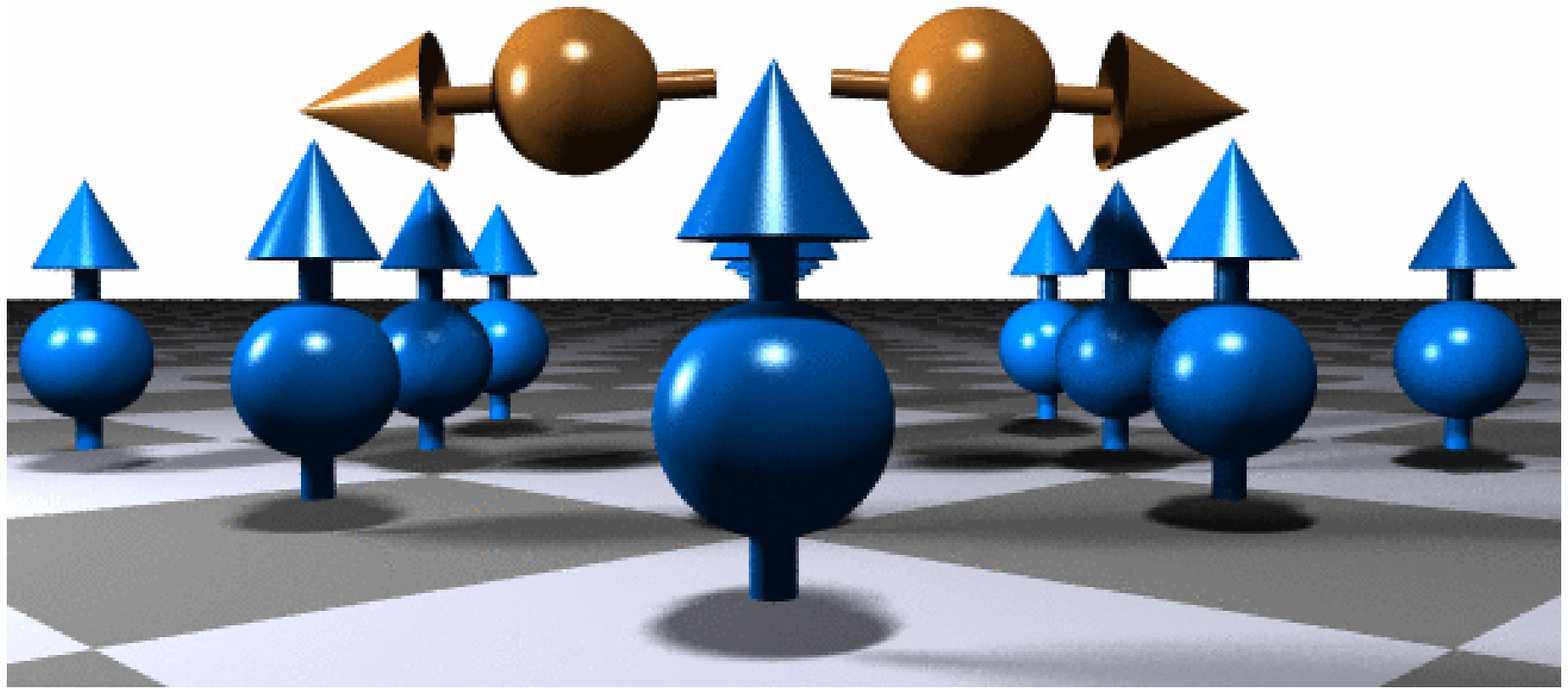}
\caption{Most stable configurations of Cr dimer-1--type obtained with (a)
the collinear KKR method and (b) the non-collinear KKR method. The
rotation angle with respect to the $z$ axis is equal to
94.2$^{\circ}$. The collinear state is the ground state, with the
non-collinear state being a local minimum (see text).}
\label{cr-dimer-ncol}
\end{center}
\end{figure}

When we allow for a rotation of the magnetic moments, non-collinear
solutions are obtained for the Cr-- and Mn--dimer-1 systems. On the
other hand the magnetic coupling of the V-- and Fe--dimer-1 remains
collinear.  Let us start with Cr--dimer-1: Fig.~\ref{cr-dimer-ncol}(a)
represents the collinear magnetic ground state. As one expects from
the adatom picture, both adatoms forming the dimer tend to couple AF
to the substrate but due to their half filled {\it d} band they also
tend to couple AF to each other. This can be understood in terms of
the Alexander--Anderson model.\cite{anderson,oswald} To give a short
explanation (see Fig.~\ref{alexander-anderson}c), both Cr and Mn have
their majority--spin VBS occupied, below $E_F$, and the minority--spin
VBS unoccupied, above $E_F$. An antiparallel configuration between
the moments in a Mn or Cr dimer lowers the energy, because the occupied
{\it d} VBS of each atom hybridizes with the unoccupied {\it d} VBS of
the other atom and is shifted to lower eigenvalues. Contrary to this,
a parallel coupling does not lower the energy, since there is no level
shifting, but only level broadening of the majority {\it
d--}VBS. Since these are fully occupied, the broadening brings no energy
gain. 

\begin{table*}
\begin{tabular}{c|rrr|rrr}
\hline
  &  &Cr & & & Mn & \\
\hline
  &Dimer 1& Dimer 2 & Dimer 3 & Dimer 1 & Dimer 2  & Dimer 3 \\
\hline
 $E_{\mathrm{FM}} - E_{\mathrm{Ferri}}$(eV) & $0.451$ &  $0.130$ & $0.120$ & $0.065$ & $-0.242$ & $-0.239$  \\
 $E_{\mathrm{AF}} - E_{\mathrm{Ferri}}$(eV) & $0.433$ & $-0.093$ & $-0.112$ & $0.496$ & $0.187$ & $0.233$\\
\hline
\end{tabular}
\caption{Energy differences between the Ferri solution and the FM (AF) configuration 
for the three types of dimers investigated.}
\label{energy-dimer}
\end{table*}

Thus there is a competition between the interatomic coupling within
the dimer, which drives it to a Ferri state, and the exchange
interaction with the substrate, which drives the moments of both atoms
in the same direction: AF for Cr and FM for Mn. As discussed in the
previous subsection, the magnetic exchange interaction (MEI) to the
substrate is relatively weak for Cr and Mn. Thus, the
\emph{intra--dimer} MEI is stronger than the MEI with the substrate,
and in the collinear approximation the ground state is found
ferrimagnetic (Ferri). Removing the collinear constraint, a compromise
can be found such that both atoms are AF coupled to each other and at
the same time (for Cr) slightly AF coupled to the substrate. This is
shown in Fig.\ref{cr-dimer-ncol}(b): the Cr adatom moments are aligned
antiparallel to each other and basically perpendicular to the
substrate moments. However, the weak AF interaction with the substrate
causes a slight tilting towards the substrate, leading to an angle of
94.2$^{\circ}$ instead of 90$^{\circ}$.  We also observe a very small
tilting ($\approx 0.3^{\circ}$) of the magnetic moments of the four
outer Ni atoms neighboring the Cr dimer (the two inner Ni atoms do not
tilt for symmetry reasons).

Despite the above considerations, the collinear Ferri state
(Fig.\ref{cr-dimer-ncol}(a)) is also a self-consistent solution of the
Kohn-Sham equations, even if the collinear constraint is
removed. Total energy calculations are needed in order to determine if
the non-collinear state is the true ground state, or if it represents a
local minimum of energy with the collinear result representing the
true ground state. After performing such calculations we find that the
ground state is collinear with an energy difference of $\Delta
E_{\mathrm{Ncol}-\mathrm{Ferri}} = 39.84$~meV (increasing the
angular-momentum cutoff to $l_{\mathrm{max}}=4$ brought no significant
change to this result).

\begin{figure}[h!]
\begin{center}
(a)
\includegraphics*[width=0.4\linewidth]{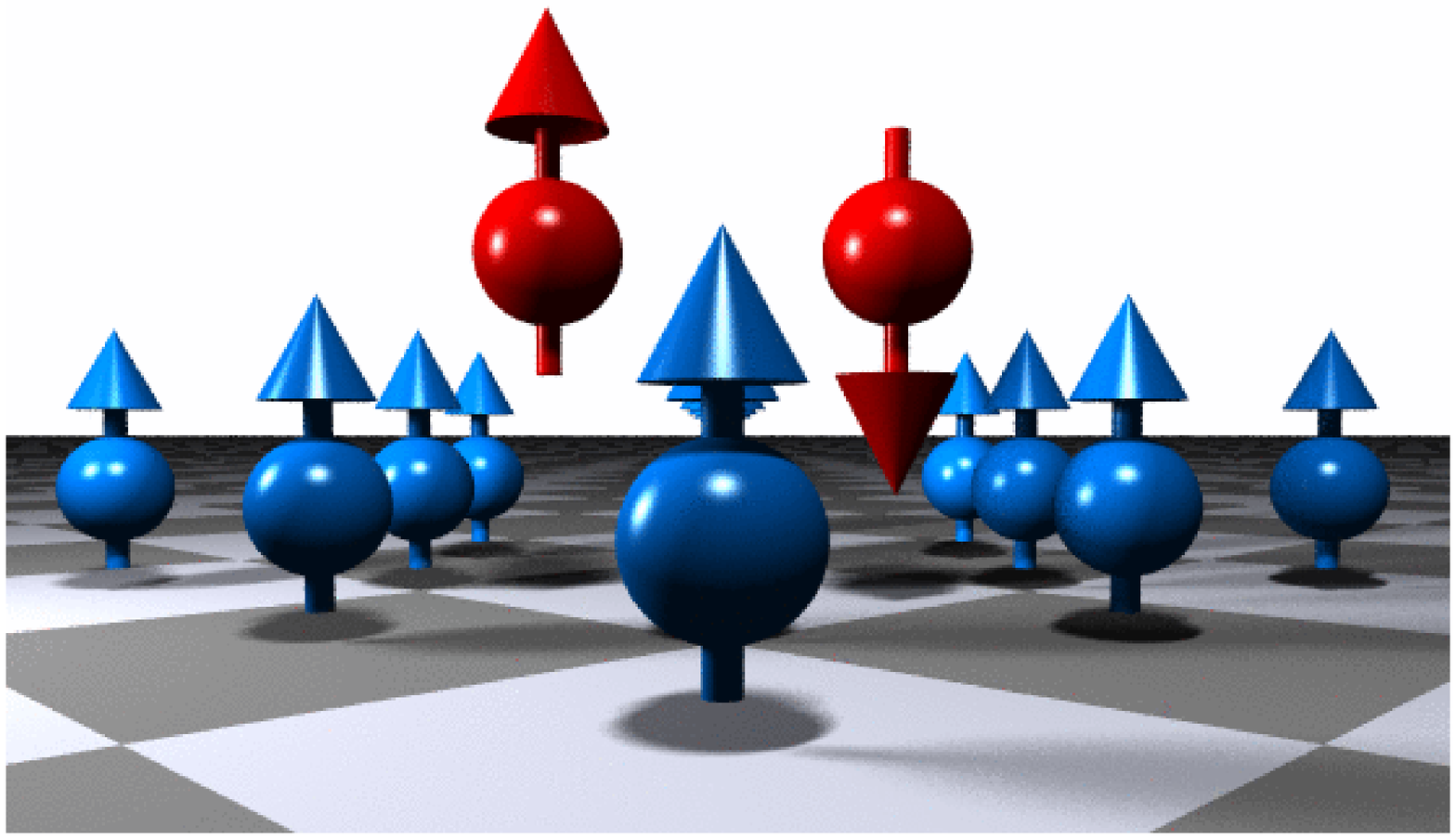}
\hspace{-0.1cm}
(b)
\includegraphics*[width=0.4\linewidth]{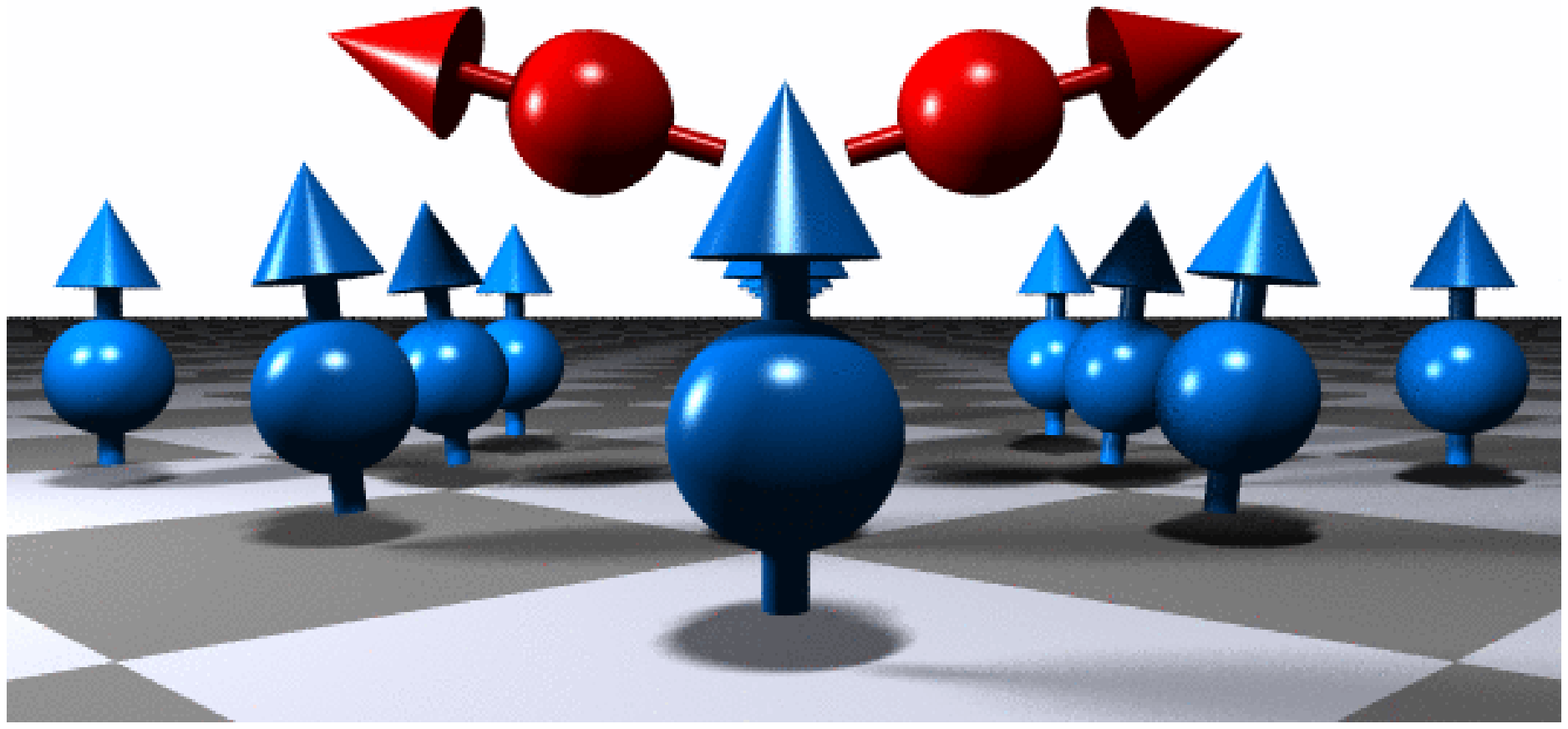}
\caption{Most stable configurations of Mn dimer-1--type obtained with
the collinear KKR method (a) and non-collinear KKR method (b). The
rotation angle with respect to the z-axis is equal to
72.6$^{\circ}$. The non-collinear state is the ground state.}
\label{mn-dimer-ncol}
\end{center}
\end{figure}
The case is different for Mn dimers. Fig.~\ref{mn-dimer-ncol} shows
the collinear and the non-collinear solutions. The dimer atoms couple
strongly antiferromagnetically to each other but the single Mn adatoms
tend to couple (weakly) ferromagnetically to the substrate. Both
adatom moments, while aligned AF with respect to each other, are
tilted in the direction of the substrate magnetization, as opposed to
the Cr--dimer. With a rotation angle of $\approx 72.6^{\circ}$, the
tilting from the $90^{\circ}$ configuration is rather large. Also the
Ni moments are tilted by $7.4^{\circ}$. The main difference with the
case of Cr--dimer-1 is that for Mn--dimer-1 the non-collinear solution
is the ground state (total energy calculations yield $\Delta
E_{\mathrm{Ncol}-\mathrm{Ferri}} = -13.45$~meV). The spin moments of
the V, Cr, Mn, and Fe dimers are given in Table~\ref{table:dimermom}.

In both cases (Cr and Mn dimers) the frustrated collinear solution is
asymmetric, while the non-collinear ground state restores the twofold
symmetry of the system. The differences in energy between the Ferri
and the non-collinear solutions are small and can be altered either by
using a different type of exchange and correlation functional such as
GGA or LSDA$+U$, or after relaxing the atoms. We note, however, that
in a test calculation we found the Cr single-adatom relaxation
to be small (3.23 \% inward with respect to the interlayer distance), and thus
we believe that the relaxation cannot affect the exchange interaction
considerably. 

As a cross-check, it is interesting to compare these non-collinear {\it
ab-initio} results to model calculations based on the Heisenberg model
with the exchange parameters fitted to the total energy
results. We assume a classical spin Hamiltonian of the form
\begin{equation}
{H} = - \frac{1}{2}\sum_{i \neq j}{J}_{ij}{\vec{e}_i \vec{e}_j}.
\label{eq:27}
\end{equation}
Here, $\vec{e}$ is a unit vector defining the direction of the
magnetic moment and $i$ and $j$ indicate the dimer atoms and their
first Ni neighbors. We can evaluate the interatomic exchange constants
$J_{\mathrm{Cr-Ni}}$, $J_{\mathrm{Mn-Ni}}$, $J_{\mathrm{Mn-Mn}}$ and
$J_{\mathrm{Cr-Cr}}$ via a fit to the total energy obtained from
collinear LSDA calculations of the FM, AF, and Ferri
configurations. Taking into account only first-neighbor interactions
and neglecting the rotation of Ni moments, we rewrite the Hamiltonian
for the dimer in terms of the tilting angles $\theta_1$ and $\theta_2$
of the two Cr (or Mn) atoms (the azimuthal angles $\phi$ do not enter
the expression because of symmetry reasons):
\begin{equation}
{H}  = - J_{\mathrm{Cr-Cr}} \cos(\theta_1 - \theta_2) 
 - 4 J_{\mathrm{Cr-Ni}} (\cos\theta_1+\cos\theta_2) +
 \mathrm{const}. 
\label{eq:28}
\end{equation}
We note the two extreme cases arising from this Heisenberg
Hamiltonian: (i) $|J_{\mathrm{Cr-Ni}}|\gg|J_{\mathrm{Cr-Cr}}|$ leads
to the stabilization of the collinear FM or AF configuration
(adatom-like behavior) and (ii)
$|J_{\mathrm{Cr-Ni}}|\ll|J_{\mathrm{Cr-Cr}}|$ leads to
antiferromagnetic coupling within the dimer if $J_{\mathrm{Cr-Cr}} < 0$.
Within the Heisenberg model the Ferri solution and the non-collinear solution 
with $\theta = 90^{\circ}$ have the same energy.

Table~{\ref{table-heisenberg1}} summarizes the estimated exchange
parameters. 
\begin{table}
\begin{tabular}{c|r|rrr}
\hline
   &   (a)  & &  (b) &\\
\hline
$J_{ij}$ (meV)& Dimer 1& Dimer 1 & Dimer 2 & Dimer 3\\
\hline
$J_{\mathrm{Cr-Ni}}$&$-1.3$&$-11.6$ &$-13.9$&$-14.5$ \\
$J_{\mathrm{Cr-Cr}}$&$-189.1$&$-221.3$&$-9.2$&$-2.0$\\
$J_{\mathrm{Mn-Ni}}$&$13.0$&$27.0$&$26.8$&$29.5$ \\
$J_{\mathrm{Mn-Mn}}$&$-138.2$&$-140.2$&$13.7$&$1.5$ \\
\hline
\end{tabular}
\caption{Values of magnetic exchange parameters $J_{ij}$ for Cr and Mn
  dimers on Ni(001), fitted from collinear first--principles total
  energy calculations (b) and obtained by the Lichtenstein
  formula\cite{lichtenstein} (a) ($J_{\mathrm{Cr-Ni}}$ and
  $J_{\mathrm{Mn-Ni}}$ are averaged over the different Ni first
  neighbours of the dimer atoms). Positive $J_{ij}$ correspond to
  ferromagnetic interactions, negative $J_{ij}$ to antiferromagnetic
  ones.}
\label{table-heisenberg1}
\end{table}
Two effects are striking: (i) The strong antiferromagnetic Cr--Cr and
Mn--Mn interaction for the dimer-1 (nearest neighbors), being an order
of magnitude larger than the exchange interactions with the substrate
and being responsible for the stabilization of the non-collinear state
structures shown in Figs.~\ref{cr-dimer-ncol} and~\ref{mn-dimer-ncol}.
(ii) The very weak Cr--Cr and Mn--Mn interactions in the dimer-2 and
-3 configurations. Whereas for the nearest-neighbors configuration
(dimer-1) the direct overlap of the {\it d}--wavefunctions of the Cr
and Mn atoms leads to the strong coupling, this overlap is missing for
larger distances and the interaction can only proceed through the
substrate. However, this interaction is weak, in fact considerably
smaller than the interaction of both adatoms with the four neighboring
Ni atoms of the substrate. Therefore these dimers are effectively
decoupled, and behave like the isolated adatoms, being
antiferromagnetically coupled to the substrate in the case of Cr and
ferromagnetically for Mn. The exchange constants $J_{ij}$ fitted to
total energy results can be compared to the ones obtained by using the
Lichtenstein formula\cite{lichtenstein} (starting from the Ferri
ground state). This rests on the force theorem, and yields the
exchange constants relevant to an infinitesimal rotation of the
moments. The results of the two methods agree best for the Mn-Mn
interaction, and reasonably well for the Cr-Cr interaction, but not
for Mn-Ni and Cr-Ni.

With the parameters from Table~{\ref{table-heisenberg1}} one can also
recalculate the non-collinear structure of the ground state. The
agreement with the {\it ab-initio} results is quite reasonable. For
the Cr dimer, one finds a slightly smaller tilting, i.e. 96$^{\circ}$
instead of 94.2$^{\circ}$, while for the Mn dimer the angle is
67.3$^{\circ}$ instead of 70.6$^{\circ}$.

The differences in energy calculated within this simple model, show
that the Cr-dimer-1 has a non-collinear ground state ($\Delta
E_{\mathrm{Ncol}-\mathrm{Ferri}} = -9.7$~meV) as well as the
Mn-dimer-1 ($\Delta E_{\mathrm{Ncol}-\mathrm{Ferri}} =
-41.6$~meV). The discrepancy obtained for the case of Cr-dimer-1 (the
LSDA calculation gives the collinear Ferri ground state) can be
attributed to the restrictions of the Heisenberg model. For instance,
for the Ferri and non-collinear configurations, the Cr moments are
slightly different, and also the reduction of the Ni moments as a
function of the rotation angle ({\it e.g.} for the single adatom)
cannot be described by the Heisenberg model, where the absolute values
of the moments are assumed to be constant.  Within the Heisenberg
model, the Ferri solution (with $\theta_1=
0^{\circ}$ and $\theta_2 = 180^{\circ}$) is
degenerate with the non-collinear solution ($\theta_{1,2} =
90^{\circ}$ with AF coupling within the dimer).

To evaluate the effect of change in coordination and hybridization, we
have undertaken a study of inatom first-neighbor dimers for V, Cr, Mn
and Fe. The V and Fe inatom-dimers were found to behave like the
adatom dimers. The V dimer is in an AF state, the Fe dimer in a FM
state, while the Cr and Mn dimers are in a Ferri state (in case of
collinear constraint). The spin moments in the collinear and
non-collinear states are given in Table~\ref{table:dimermom}
\begin{table*}
\begin{tabular}{lcccccc}
\hline
Dimer type & V(AF)         & Cr(Ferri)    &  Cr(Ncol)   &Mn(Ferri)     & Mn(Ncol)    & Fe(FM) \\
\hline
On Ni(001) &($-1.28,-1.28$)&($-3.04,3.05$)&($3.03,3.03$)&($-3.84,3.69$)&($3.75,3.75$)&($3.10,3.10$)\\
In Ni(001) &($-0.32,-0.32$)&($-2.00,1.96$)&($1.97,1.97$)&($-3.32,3.20$)&($3.26,3.26$)&($2.88,2.88$)\\
\hline
\end{tabular}
\caption{Atomic spin moments (in $\mu_B$) of the adatom and inatom
  dimers (of type 1, {\it i.e.}, nearest-neighbors) in the collinear
  and non-collinear configurations. A minus sign of the collinear
  moments indicates an antiparallel orientation with respect to the
  substrate magnetization. Embedding the dimer into the surface
  causes, as expected, a decrease of the spin moments due to stronger
  hybridization of the $d$ wavefunctions.\label{table:dimermom}}
\end{table*}
Within the Ferri-dimers, the difference between the moments of the two
atoms is due to the different kind of coupling that each inatom has
with the the substrate (AF or FM). One notices also that the magnetic
moments in the ground state decrease compared to the values obtained
for the single inatoms and single adatoms. When the rotation of the
moments is allowed, Cr--dimer can be stabilized at an angle of
107$^{\circ}$ (instead of 94.2$^{\circ}$ found for the adatom--dimer
case), and Mn--dimer at an angle of 80.9$^{\circ}$ (instead of
72.6$^{\circ}$). Thus the non-collinear solutions obtained for
inatom--dimers are rather similar to what was obtained for
adatom--dimers.  Energetically, however, both the Cr and the Mn inatom
dimers show a lower total energy in the collinear Ferri state (for Cr,
$\Delta E_{\mathrm{Ncol}-\mathrm{Ferri}} = 24.11$~meV; for Mn, $\Delta
E_{\mathrm{Ncol}-\mathrm{Ferri}} = 22.5$~meV).

\subsection{Trimers}

\begin{figure}[h!]
\begin{center}
(a)
\includegraphics*[width=0.35\linewidth]{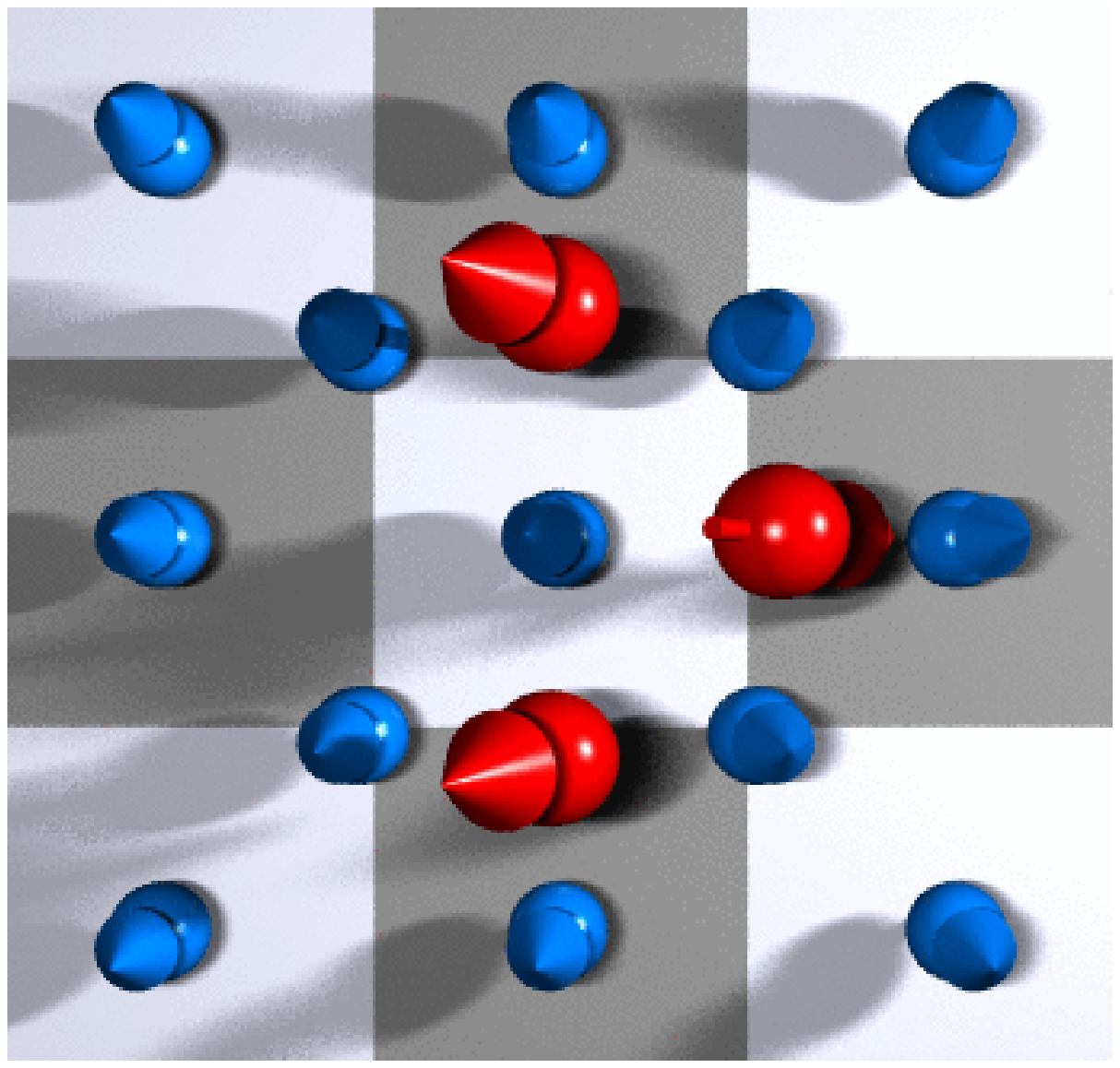}
(b)
\includegraphics*[width=0.45\linewidth]{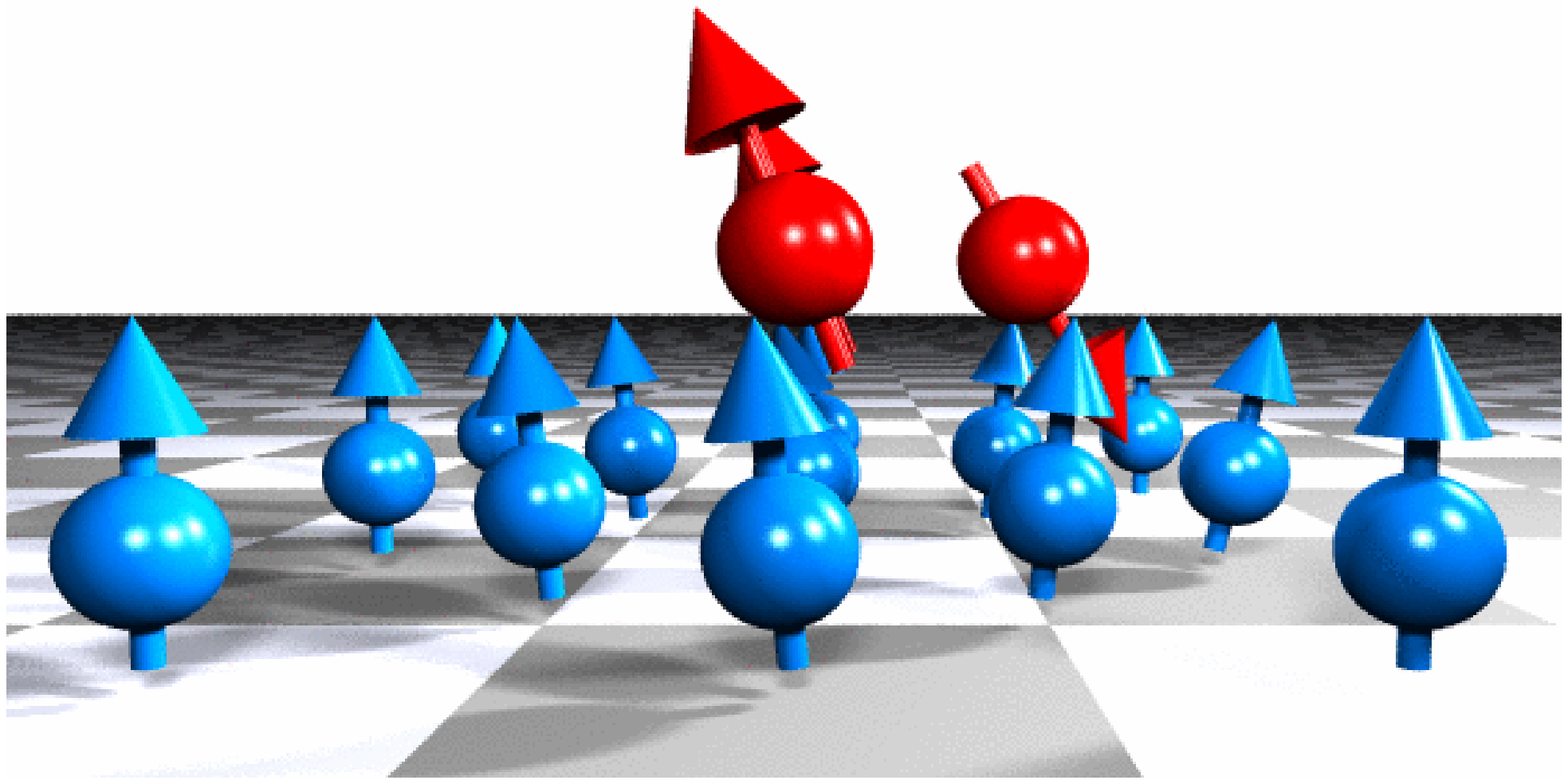}
\caption{Non-collinear state of the Mn trimer on Ni(001) surface. Side
view (a) and front view (b) are shown. This represents a local minimum
in energy, with the collinear state being the ground state (see
text).}
\label{mn-trimer-front}
\end{center}
\end{figure}

Following the same procedure as for dimers we first investigated
several collinear magnetic configurations for the most compact trimer
on the Ni(001) surface, which has the shape of an isosceles
rectangular triangle (see Fig.~\ref{mn-trimer-front}) of side
$\sqrt{2}a/2$ and hypotenuse $a$ ($a$ is the Ni fcc lattice
constant). It is expected, and verified by total-energy calculations,
to find the $\downarrow\uparrow\downarrow$ configuration as the
collinear magnetic ground state for Cr and the
$\uparrow\downarrow\uparrow$ for the Mn trimer ($\uparrow$ means an
atomic moment parallel to the substrate, $\downarrow$ an antiparallel
one; the middle arrow represents the direction of the atomic moment at
the right-angle corner of the triangle). In Table~\ref{energy-trimer},
the energy differences among the possible collinear configurations are
given; for the $\uparrow \uparrow \uparrow$ and $\downarrow
\downarrow \downarrow$ Cr trimers, no self-consistent solution could
be found.
\begin{table}
\begin{tabular}{ccccrcc}
\hline
Magn. Config. & $\uparrow \uparrow \downarrow$ & $\downarrow \uparrow \downarrow$ & 
$\downarrow \downarrow \uparrow $& $\uparrow \downarrow \uparrow$ & 
 $\uparrow \uparrow \uparrow$ & $\downarrow \downarrow \downarrow$\\
\hline
Cr: $E - E_{\downarrow \uparrow \downarrow}$(eV) & 0.420 &  0 & 0.390 & 0.193 & --- & ---  \\
Mn: $E - E_{\downarrow \uparrow \downarrow}$(eV) & 0.116 & 0 & 0.318 &  $-0.184$  & 0.239 & 0.817\\
\hline
\end{tabular}
\caption{Energy differences between the different calculated collinear
magnetic configurations with the $\downarrow \uparrow \downarrow$
configuration. The direction of the arrows represents the direction of
the atomic moments relative to the substrate magnetization ($\uparrow$
parallel, $\downarrow$ antiparallel). The middle arrow represents the
atom at the right-angle corner of the trimer.}
\label{energy-trimer}
\end{table}

Allowing free rotation of the magnetic moments leads to no change for
the Cr trimer $\downarrow\uparrow\downarrow$---the state remains
collinear (within numerical accuracy). On the other hand, for the Mn
trimer a non-collinear solution is found (Fig.~\ref{mn-trimer-front})
with the nearest neighbours almost antiferromagnetic to each other,
but with a collective tilting angle with respect to the
substrate. This tilting angle is induced by the ferromagnetic MEI
between the central Mn atom with the substrate, competing with the
antiferromagnetic MEI with its two companions. The top view of the
surface shows that the in-plane components of the magnetic moments are
collinear.

The tilting is somewhat smaller ($21.7^{\circ}$) for the two Mn atoms
with moments up than for the Mn atom with moment down
($28.5^{\circ}$). Also the neighboring Ni--surface atoms experience
small tilting, with varying angles around $4^{\circ}-10^{\circ}$.
From the energy point of view, the ground state is the collinear one,
$\uparrow \downarrow \uparrow$, with an energy difference of $\Delta
E_{\mathrm{Ncol}-\uparrow \downarrow \uparrow} = 22.92$~meV with
respect to the non-collinear solution. 

We have also investigated the cases where the trimers are sitting in
the surface layer. No non-collinear solution was found, while there is
no change in the collinear ground state which is
$\downarrow\uparrow\downarrow$ for the Cr trimer and $\uparrow
\downarrow \uparrow$ for Mn trimer.

One should note that the moments of the two first neighboring 
impurities are almost compensated in the Ferri solution. The third 
moment determines the total interaction between the substrate 
and the trimer which has then a net moment 
coming mainly from the additional 
impurity. This interaction is identical to the single adatom (or inatom) 
type of coupling.

\section{Summary}

We have presented a formalism for the treatment of non-collinear
magnetic clusters on surfaces and in bulk, based on the Green function
technique of Korringa, Kohn and Rostoker, and on spin density
functional theory. We have applied the formalism on the study of small
transition metal clusters (dimers and trimers) on and in the Ni(001)
surface.

Emphasis was placed on Cr and Mn clusters, for which we found that
magnetic frustration can lead to non-collinear magnetic order. The
origin of the frustration is the competition of the antiferromagnetic
exchange coupling among the Cr or Mn atoms with the antiferromagnetic
(for Cr) or ferromagnetic (for Mn) exchange coupling between the
adatoms and the substrate. In this respect, the result is different
than the prototype non-collinear configurations arising from
antiferromagnetic interactions among atoms in triangular geometry.

We found that Cr and Mn first-neighbouring adatom dimers can show
non-collinear behavior, while increasing the distance between the
adatoms of the dimer leads to the same state as for single adatoms.
The energy differences between the collinear ferrimagnetic state and
the non-collinear one are $\Delta E^{\mathrm{Cr}}_{\mathrm{Ncol-Ferri}}
= 39.84$~meV (the ground state is collinear), $\Delta
E^{\mathrm{Mn}}_{\mathrm{Ncol-Ferri}}= -13.45$~meV (the ground state
is non-collinear). Embedding the dimers in the first surface layer
restores the Ferri collinear solution as a ground state also for Mn
adatom dimers ( $\Delta E^{\mathrm{Cr}}_{\mathrm{Ncol-Ferri}} =
24.11$~meV, $\Delta E^{\mathrm{Mn}}_{\mathrm{Ncol-Ferri}}=
22.5$~meV). 

Our {\it ab-initio} results for dimers are compared to the solution of
a classical Heisenberg model with exchange parameters fitted to total
energy calculations. The agreement for the tilting angles in the
non-collinear state is good, but the Heisenberg model does not capture
the collinear ground state for the Cr dimer. This discrepancy occurs
because the Heisenberg model is restricted to constant absolute values
of localized spins.

The trimers studied so far are characterized by a collinear ground
state: $\downarrow \uparrow \downarrow$ for the Cr trimer and
$\uparrow \downarrow \uparrow$ for the Mn trimer. The Mn trimer has
also a non-collinear metastable solution with an energy difference
$\Delta E_{\mathrm{Ncol}-\uparrow \downarrow \uparrow} = 22.92$~meV.

We believe that the energetic proximity of the collinear to the
non-collinear states is directly related to the weakness of the
exchange interaction with the Ni substrate. Replacing it by an fcc Fe
substrate will possibly change the ground state drastically.  Work in
this direction progress and will be reported elsewhere.

\section{acknowledgments}
We would like to thank Rudi Zeller for fruitful discussions.
This work was financed by the Priority Program ``Clusters in Contact
with Surfaces'' (SPP 1153) of the Deutsche Forschungsgemeinschaft.

\end{document}